\documentclass[iop,apj]{emulateapj}
\usepackage{graphicx}
\usepackage{epstopdf}
\usepackage{amsmath}
\usepackage{amsfonts}
\usepackage{amssymb}
\usepackage[usenames, dvipsnames]{color}
\usepackage{hyperref}

\newcommand{\spreadmodel}{\ensuremath{spread\_model~}}

\begin{document}

\title{The Phoenix stream: a cold stream in the Southern hemisphere}

\author{
    E.~Balbinot\altaffilmark{1},
    B.~Yanny\altaffilmark{2},
    T.~S.~Li\altaffilmark{3},
    B.~Santiago\altaffilmark{4,5},
    J.~L.~Marshall\altaffilmark{3},
    D.~A.~Finley\altaffilmark{2},
    A. Pieres\altaffilmark{4,5},
    T. M. C.~Abbott\altaffilmark{6},
    F.~B.~Abdalla\altaffilmark{7},
    S.~Allam\altaffilmark{2},
    A.~Benoit-L{\'e}vy\altaffilmark{7},
    G.~M.~Bernstein\altaffilmark{8},
    E.~Bertin\altaffilmark{9,10},
    D.~Brooks\altaffilmark{7},
    D.~L.~Burke\altaffilmark{11,12},
    A.~Carnero~Rosell\altaffilmark{5,13},
    M.~Carrasco~Kind\altaffilmark{14,15},
    J.~Carretero\altaffilmark{16,17},
    C.~E.~Cunha\altaffilmark{11},
    L.~N.~da Costa\altaffilmark{5,13},
    D.~L.~DePoy\altaffilmark{3},
    S.~Desai\altaffilmark{18,19},
    H.~T.~Diehl\altaffilmark{2},
    P.~Doel\altaffilmark{7},
    J.~Estrada\altaffilmark{2},
    B.~Flaugher\altaffilmark{2},
    J.~Frieman\altaffilmark{2,20},
    D.~W.~Gerdes\altaffilmark{21},
    D.~Gruen\altaffilmark{22,23},
    R.~A.~Gruendl\altaffilmark{14,15},
    K.~Honscheid\altaffilmark{24,25},
    D.~J.~James\altaffilmark{6},
    K.~Kuehn\altaffilmark{26},
    N.~Kuropatkin\altaffilmark{2},
    O.~Lahav\altaffilmark{7},
    M.~March\altaffilmark{8},
    P.~Martini\altaffilmark{24,27},
    R.~Miquel\altaffilmark{28,17},
    R.~C.~Nichol\altaffilmark{29},
    R.~Ogando\altaffilmark{5,13},
    A.~K.~Romer\altaffilmark{30},
    E.~Sanchez\altaffilmark{31},
    M.~Schubnell\altaffilmark{21},
    I.~Sevilla-Noarbe\altaffilmark{31,14},
    R.~C.~Smith\altaffilmark{6},
    M.~Soares-Santos\altaffilmark{2},
    F.~Sobreira\altaffilmark{2,5},
    E.~Suchyta\altaffilmark{24,25},
    G.~Tarle\altaffilmark{21},
    D.~Thomas\altaffilmark{29},
    D.~Tucker\altaffilmark{2},
    A.~R.~Walker\altaffilmark{6}
    \\ \vspace{0.2cm} (The DES Collaboration) \\
}
 
\altaffiltext{1}{Department of Physics, University of Surrey, Guildford GU2 7XH, UK}
\altaffiltext{2}{Fermi National Accelerator Laboratory, P. O. Box 500, Batavia, IL 60510, USA}
\altaffiltext{3}{George P. and Cynthia Woods Mitchell Institute for Fundamental Physics and Astronomy, and Department of Physics and Astronomy, Texas A\&M University, College Station, TX 77843,  USA}
\altaffiltext{4}{Instituto de F\'\i sica, UFRGS, Caixa Postal 15051, Porto Alegre, RS - 91501-970, Brazil}
\altaffiltext{5}{Laborat\'orio Interinstitucional de e-Astronomia - LIneA, Rua Gal. Jos\'e Cristino 77, Rio de Janeiro, RJ - 20921-400, Brazil}
\altaffiltext{6}{Cerro Tololo Inter-American Observatory, National Optical Astronomy Observatory, Casilla 603, La Serena, Chile}
\altaffiltext{7}{Department of Physics \& Astronomy, University College London, Gower Street, London, WC1E 6BT, UK}
\altaffiltext{8}{Department of Physics and Astronomy, University of Pennsylvania, Philadelphia, PA 19104, USA}
\altaffiltext{9}{CNRS, UMR 7095, Institut d'Astrophysique de Paris, F-75014, Paris, France}
\altaffiltext{10}{Sorbonne Universit\'es, UPMC Univ Paris 06, UMR 7095, Institut d'Astrophysique de Paris, F-75014, Paris, France}
\altaffiltext{11}{Kavli Institute for Particle Astrophysics \& Cosmology, P. O. Box 2450, Stanford University, Stanford, CA 94305, USA}
\altaffiltext{12}{SLAC National Accelerator Laboratory, Menlo Park, CA 94025, USA}
\altaffiltext{13}{Observat\'orio Nacional, Rua Gal. Jos\'e Cristino 77, Rio de Janeiro, RJ - 20921-400, Brazil}
\altaffiltext{14}{Department of Astronomy, University of Illinois, 1002 W. Green Street, Urbana, IL 61801, USA}
\altaffiltext{15}{National Center for Supercomputing Applications, 1205 West Clark St., Urbana, IL 61801, USA}
\altaffiltext{16}{Institut de Ci\`encies de l'Espai, IEEC-CSIC, Campus UAB, Carrer de Can Magrans, s/n,  08193 Bellaterra, Barcelona, Spain}
\altaffiltext{17}{Institut de F\'{\i}sica d'Altes Energies, Universitat Aut\`onoma de Barcelona, E-08193 Bellaterra, Barcelona, Spain}
\altaffiltext{18}{Excellence Cluster Universe, Boltzmannstr.\ 2, 85748 Garching, Germany}
\altaffiltext{19}{Faculty of Physics, Ludwig-Maximilians University, Scheinerstr. 1, 81679 Munich, Germany}
\altaffiltext{20}{Kavli Institute for Cosmological Physics, University of Chicago, Chicago, IL 60637, USA}
\altaffiltext{21}{Department of Physics, University of Michigan, Ann Arbor, MI 48109, USA}
\altaffiltext{22}{Max Planck Institute for Extraterrestrial Physics, Giessenbachstrasse, 85748 Garching, Germany}
\altaffiltext{23}{Universit\"ats-Sternwarte, Fakult\"at f\"ur Physik, Ludwig-Maximilians Universit\"at M\"unchen, Scheinerstr. 1, 81679 M\"unchen, Germany}
\altaffiltext{24}{Center for Cosmology and Astro-Particle Physics, The Ohio State University, Columbus, OH 43210, USA}
\altaffiltext{25}{Department of Physics, The Ohio State University, Columbus, OH 43210, USA}
\altaffiltext{26}{Australian Astronomical Observatory, North Ryde, NSW 2113, Australia}
\altaffiltext{27}{Department of Astronomy, The Ohio State University, Columbus, OH 43210, USA}
\altaffiltext{28}{Instituci\'o Catalana de Recerca i Estudis Avan\c{c}ats, E-08010 Barcelona, Spain}
\altaffiltext{29}{Institute of Cosmology \& Gravitation, University of Portsmouth, Portsmouth, PO1 3FX, UK}
\altaffiltext{30}{Department of Physics and Astronomy, Pevensey Building, University of Sussex, Brighton, BN1 9QH, UK}
\altaffiltext{31}{Centro de Investigaciones Energ\'eticas, Medioambientales y Tecnol\'ogicas (CIEMAT), Madrid, Spain}

\email{Correspondece to: e.balbinot@surrey.ac.uk}

\shorttitle{The Phoenix stream}
\shortauthors{E. Balbinot et al.}

\begin{abstract}
We report the discovery of a stellar stream in the Dark Energy Survey (DES) Year
1 (Y1A1) data. The discovery was made through simple color-magnitude filters and
visual inspection of the Y1A1 data. We refer to this new object as the Phoenix
stream, after its resident constellation. After subtraction of the
background stellar population we detect a clear signal of a simple stellar
population.  By fitting the ridge line of the stream in color-magnitude space,
we find that a stellar population with age $\tau=11.5\pm0.5$ Gyr and
$[Fe/H]<−1.6$ located 17.5$\pm$0.9 kpc from the Sun gives an adequate
description of the stream stellar population. The stream is detected over an
extension of 8$\fdg$1 (2.5 kpc) and has a width of $\sim$54 pc assuming a
Gaussian profile, indicating that a globular cluster is a probable progenitor.
There is no known globular cluster within 5 kpc compatible with being the
progenitor of the stream, assuming that the stream traces its orbit. We examined
overdensities along the stream, however no obvious counterpart bound stellar
system is visible in the coadded images. We also find overdensities along the
stream that appear to be symmetrically distributed - consistent with the
epicyclic overdensity scenario for the formation of cold streams - as well as a
misalignment between the Northern and Southern part of stream. Despite
the close proximity we find no evidence that this stream and the halo
cluster NGC 1261 have a common accretion origin linked to the recently found
EriPhe overdensity \citep{eriphe}.
\end{abstract}
\keywords{Galaxy: halo -- Galaxy: structure}

\maketitle

\section{Introduction}

Our understanding of the structure of the Galactic halo has evolved considerably
in the past two decades, largely thanks to deep and homogeneous photometric
surveys, such as the Two Micron All Sky Survey \citep[2MASS;][]{2MASS} and the
Sloan Digital Sky Survey \citep[SDSS;][]{SDSS}.  The stellar halo is now known
to be inhabited by a variety of spatial and kinematic stellar substructure, from
globular clusters (GCs) and dwarf galaxies to extended stellar clouds and
streams \citep[see e.g.][]{2005ApJ...626L..85W, 2006ApJ...647L.111B,
2007ApJ...654..897B}. In fact, recent simulations based on hierarchical models
of structure formation predict that most halo stars were brought by the
disruption of the Galactic substructures \citep{2001ApJ...548...33B}.

The thin and cold stellar streams found in the Galaxy often span tens of
degrees on the sky and originate from the tidal effects of the host on the
progenitor, whether a GC or a dwarf galaxy. Perhaps the most conspicuous
examples of Galactic streams are those associated with the Pal 5 GC and the
Sagittarius dwarf \citep{2001ApJ...548L.165O, 2002ApJ...569..245N}. The tidal
nature of such streams makes them useful probes of the dark matter distribution
across the halo \citep{2005ApJ...619..800J, Kuepper15}. Detailed modeling of a
stream's position, distance, kinematics, gaps and overdensities in extended cold
streams also leads to constraints on the amount of dark matter fragments
orbiting the halo, known as subhalos \citep{2011ApJ...731...58Y,
2015ApJ...803...75N}, on the progenitor's properties and on Galactic parameters
\citep{2010ApJ...712..260K}.

The Dark Energy Survey \citep[DES;][]{Abbott:2005bi} is an on-going deep ($g
\sim 24.7$) photometric survey in the Southern hemisphere that started its
planned 5-year mission of collecting data in 2013. Despite its focus on
cosmology, DES data have already produced a wealth of results pertaining to
resolved stellar populations in the Galaxy and its vicinity, including the
analysis of the structure and stellar populations in the outskirts of the Large
Magellanic Cloud \citep{Balbinot15}, the identification of new Galactic
companions \citep{Bechtol15, Koposov15, Luque15, DESsats15b, Kim15}, and
the development of a new search for variable stars exclusively based on DES
data (Hatt et al., in preparation).

We here report on the discovery of the first cold stellar stream using DES
data. In Section 2 we give more details about DES, the data used, and the
search algorithm. Our results are presented in Section 3 and our conclusions
are in Section 4.

\section{Data Analysis}

DES is a wide-field optical imaging survey using broad photometric bands
($grizY$) performed with the Dark Energy Camera \citep[DECam; described in
detail in][]{Flaugher15}. The DECam focal plane is comprised of 74 CCDs: 62
2k$\times$4k CCDs dedicated to science imaging and 12 2k$\times$2k CCDs for
guiding, focus, and alignment.  DECam is installed at the prime focus of the
4-meter Blanco telescope at Cerro Tololo Inter-American Observatory. In this
configuration, DECam has a 2.2-degree-wide field-of-view and a central pixel
scale of 0.263 arcseconds. The full DES survey is scheduled for 525 nights
distributed over five years. Here, we consider data from the first year of DES
obtained between 15 August 2013 to 9 February 2014.

The first internal annual release of DES data (Y1A1) comprises the data products
obtained from the processing of a subset of wide- and supernova- field data
accumulated during the first year of DES operations \citep{Y1paper}.  Briefly,
the image processing pipeline consists of image detrending (crosstalk
correction, bias subtraction, flat-fielding, etc), astrometric calibration,
nightly photometric calibration, global calibration, image coaddition, and
object catalog creation.  For a more detailed description of the DESDM image
processing pipeline, we refer to \citet{2012ApJ...757...83D,
2012SPIE.8451E..0DM} and for a recent overview see \citet{Balbinot15}. The
\textsc{SExtractor} toolkit is used to create image catalogs from the processed
and coadded images \citep{2011ASPC..442..435B, 1996A&AS..117..393B}. The number
of overlapping exposures in Y1A1 varies, but most of the footprint has at
least 3 coadded exposures. The Y1A1 coadd object catalog contains $\sim 131$
million unique objects spread over $\sim1{,}800 \deg^2$.  This area includes
$\sim 200$ deg$^2$ overlapping with the Stripe-82 region of SDSS, as well as a
contiguous region of $\sim 1{,}600\deg^2$ overlapping the South Pole Telescope
(SPT) footprint.

We perform stellar selection on the Y1A1 coadd object catalog based on the
\spreadmodel quantity output of \textsc{SExtractor}~\citep{2012ApJ...757...83D}.
To avoid issues arising from fitting the PSF across variable depth coadded
images, we utilize the weighted-average (wavg) of the \spreadmodel measurements
from the single-epoch exposures. Our stellar sample consists of well-measured
objects with $|wavg\_spread\_model\_i| \allowbreak < \allowbreak 0.003$,
$flags_{\{g,r,i\}} < 4$, and $magerr\_auto\_\{g,r,i\} < 1$ (henceforth referred
to as ``stars''). Our stellar completeness is ${>}\,90\%$ down to magnitude $g
\sim 22$, at which point it drops to $\sim 50\%$ by $g \sim 23$
\citep{Bechtol15}.

Stars are extinction corrected according to \citet{SFD} with the scaling
correction from \citet{2011ApJ...737..103S} assuming the extinction curve from
\citet{Cardelli89} and a calibration at infinity, that is we assume that the
light of every object in our sample crosses the full extent of the dust column
measured in the dust maps.

From this point all magnitudes considered in this paper are corrected
for the extinction.

\subsection{Search method}

\begin{figure*}
\centering
\includegraphics[width=0.85\textwidth]{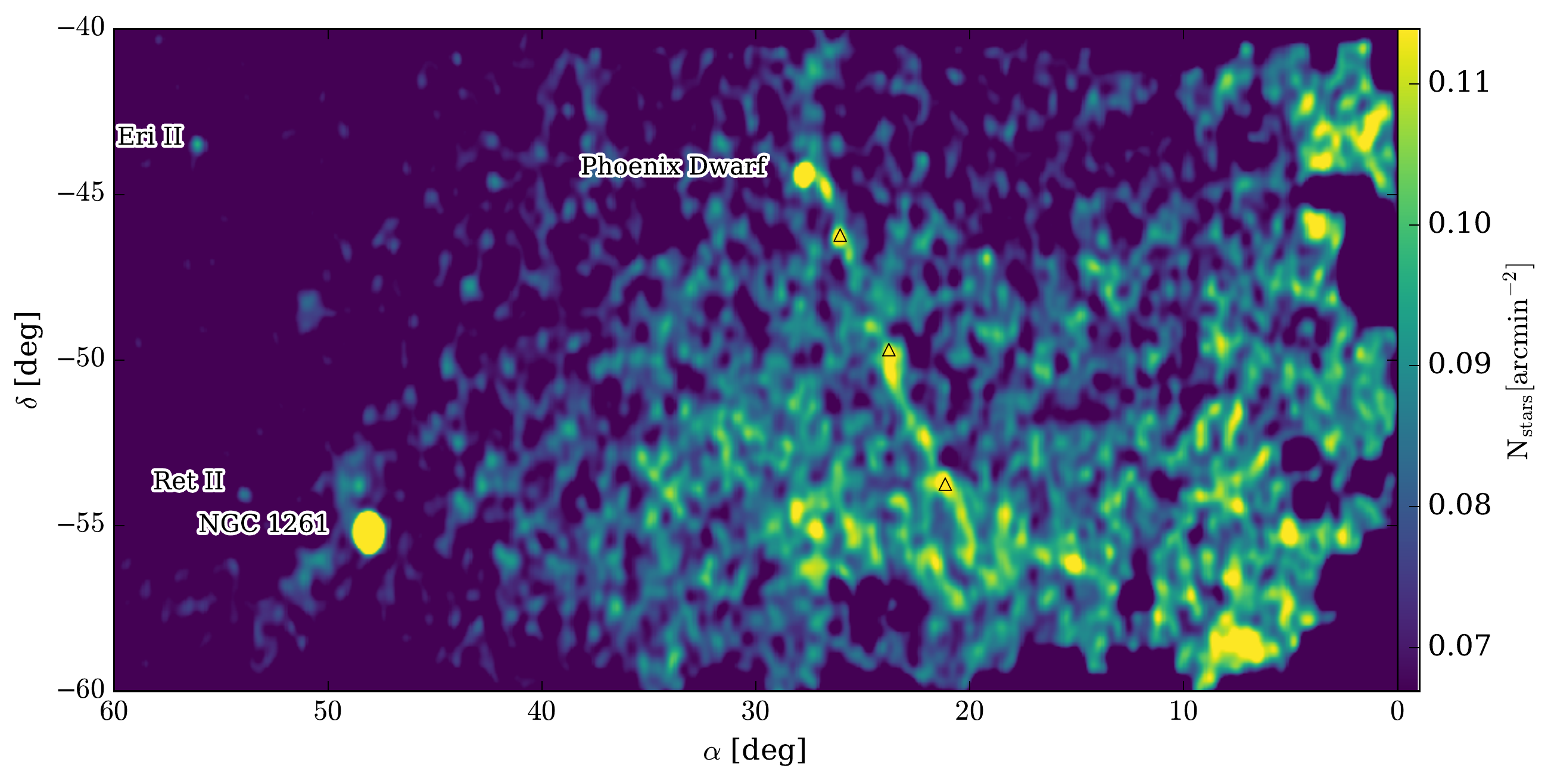}
  \caption{Y1A1 density map for stars with $0.2 < (g-r) < 0.6$ and $20 < g <
      23$. The open triangles show the anchor points adopted for the stream.
      Other interesting objects are labeled in the figure. This density map was
      convolved with a 2$\times$1-pixel Gaussian. Each pixel has a size of
  $4.5\arcmin\times2.8\arcmin$.}
\label{colorselect}
\end{figure*}

Using the objects classified as stars according to the criteria described in the
previous section, we apply narrow color filters to isolate interesting stellar
types such as old turnoff stars and horizontal branch (HB) stars. To avoid
issues related to the inhomogeneous photometric depth of the survey and
saturation of bright stars, we only use stars with magnitudes $17 < g < 23$. We
find that this magnitude limit yields a sample that has a completeness that is
fairly constant across the footprint and produces a smoothly varying density for
the field stars.

For each color-selected catalog we build a sky ``density map". Throughout this
paper we use density maps to refer to maps where we show the number of sources
per pixel (N) in a Cartesian projection. The pixel size is made explicit
whenever necessary. The pixel area is corrected for changes in solid angle
with declination. These density maps are visually inspected for overdensities.
In \autoref{colorselect} we show the particularly interesting density map for
stars with $0.2 < (g-r) < 0.6$ and $20 < g < 23$, which selects mainly turnoff
and upper main-sequence stars from an old ($>$ 10 Gyr) simple stellar population
(SSP), according to \citet{2012MNRAS.427..127B}. Several features are noticeable,
such as the globular cluster NGC 1261, the Phoenix dwarf galaxy
\citep{Canterna77}, and two of the recently discovered dwarf galaxies
\citep{Bechtol15, Koposov15}: Reticulum II (Ret II) and Eridanus II (Eri II).
These objects are labeled with their names in the figure. A linear
structure is also visible near the Phoenix dwarf extending from $(\alpha,
\delta) \simeq (20^{\circ}, -57^{\circ})$ to $(27^{\circ}, -45^{\circ})$. This
structure is highlighted by open triangles marking high density points along the
stream candidate.

In the same figure, a large overdensity of stars is visible between the stream
candidate and NGC 1261. This feature is the Eridanus-Phoenix (EriPhe)
overdensity and it is discussed in detail in a simultaneous publication \citep{eriphe}. 

\section{Results}
\label{sec:res}

By means of the method outlined in the previous section we perform a
visual search for stellar overdensities. This search, conducted over
the full Y1A1 footprint, has revealed only one\footnotemark[1] stream candidate
which is shown in \autoref{colorselect}. For simplicity, we refer to this
candidate stream as the Phoenix stream due to its proximity to the Phoenix
constellation.

\footnotetext[1]{Recently \citet{Mackey15} reported a 10 kpc long
stream associated to the Large Magellanic Cloud (LMC) using a similar data set
to the one in this work. We confirm this detection, however its tidal origin is
still uncertain.}

\begin{figure}
\centering
  \includegraphics[width=0.45\textwidth]{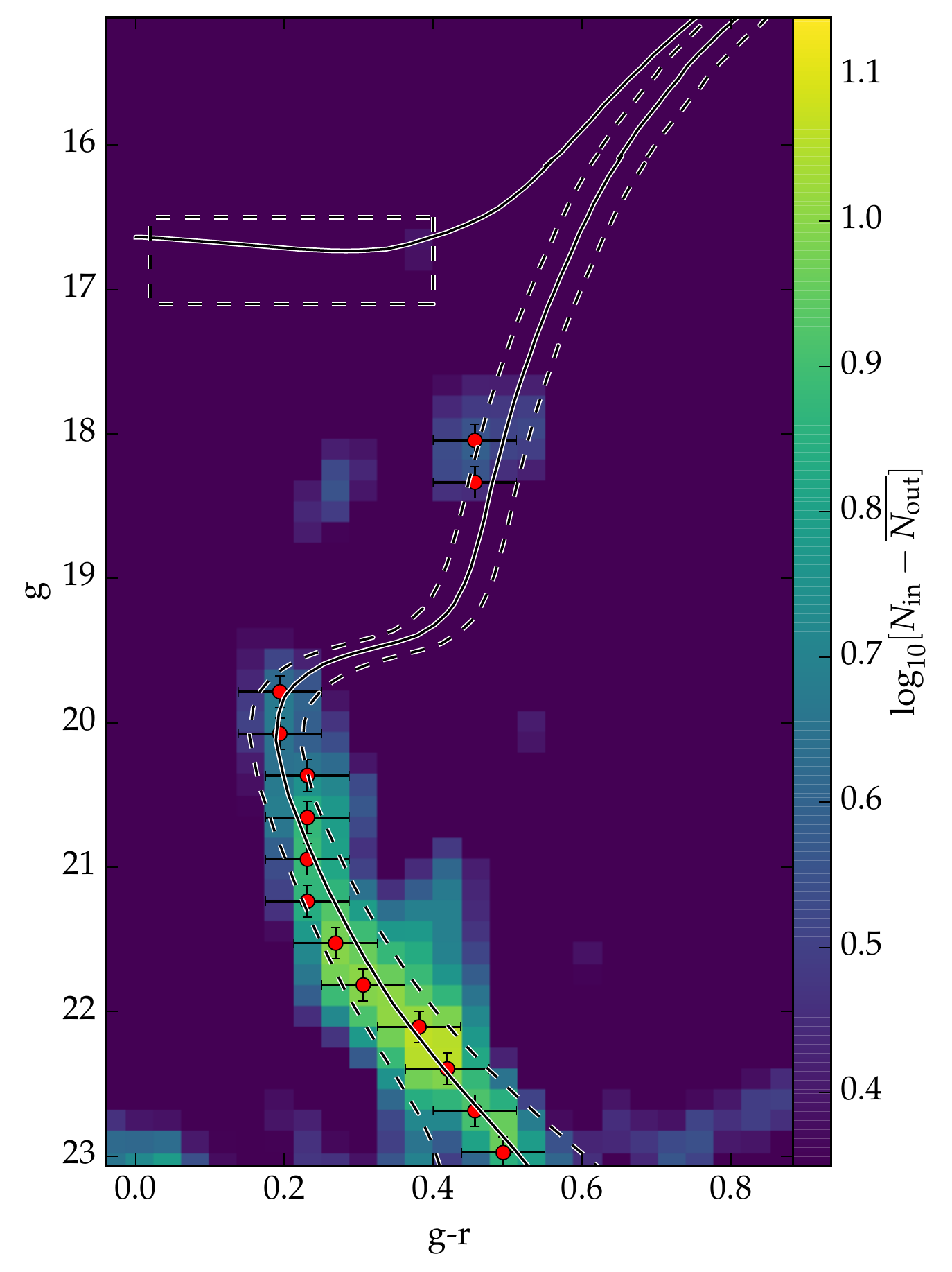} 
  \caption{Decontaminated Hess diagram of the stream candidate. The
      decontamination process is described in detail in the text. The solid line
      shows an 11.5 Gyr and ${\rm [Fe/H]}$=-1.9 PARSEC isochrone
      \citep{2012MNRAS.427..127B}. The dashed lines show the color magnitude
  diagram (CMD) region selected to isolate stream stars, including a box to
  select Horizontal Branch (HB) stars. The red dots with error bars form the
  ridge line used to perform the isochrone fit.}
  \label{cmd_hess}
\end{figure}

To study the stellar population that comprises the Phoenix stream we
define a line passing through the center of the stream using anchor points along
the stream (three open triangles in \autoref{colorselect}). We then select stars
inside a box defined by the stream central line with an offset of $\pm$
0.8$^{\circ}$ in RA. We name this selection \emph{on stream}. To compare with
the typical Milky Way (MW) stellar population at this position in the sky we
select stars in boxes that are offset by $\pm 1.5^{\circ}$ with respect to the
central one and have the same width as the \emph{on stream} region. We name
these selections \emph{off stream east} and \emph{west}. For each region
described above we compute the solid angle normalized Hess diagram.
We use \textsc{Mangle} masks \citep{Swanson08} to compute the solid
angle of each box taking into account possible holes in the survey footprint.
In \autoref{cmd_hess} we show the Hess diagram of the \emph{on stream} minus the
average diagram of the two \emph{off stream} ones in logarithmic scale.

From the decontaminated Hess diagram shown in \autoref{cmd_hess} we estimate
that the stream has $\sim$ 500 stars that fall within the photometric limits of
DES. This decontaminated Hess diagram was smoothed using a Gaussian kernel with
a dispersion of 0.06$\times$0.2 in color and magnitude respectively. This step
is required due to the low number of stars in the stream and allows us to define
a ridge line shown as the red circles in \autoref{cmd_hess}. The ridge line is
defined as the peak value of counts in color for each magnitude bin, the bin
size for the ridge line construction being twice as large as the one used for
the Hess diagram. Magnitude bins with low counts or peak values that obviously
depart from the bulk of the stream stars are discarded. We define the error bars
as equal to the Gaussian kernel size, which is always larger than the photometric
errors in the magnitude range shown. This choice of error bar accounts for the
broadening of the Hess diagram due to the smoothing process. Using the typical
photometric error would yield unrealistically small uncertainty estimates that
would propagate into the stellar evolution model fitting, described below.

We compare the ridge line to different \textsc{PARSEC} stellar evolution models
\citep{2012MNRAS.427..127B} by computing the minimum distance from each ridge
line point to a given model. The model grid has a resolution of 0.01 in
$\log_{10}(\tau/yr)$ in the range from 9 to 10.16 and 0.0002 in Z in the range
from 0.0001 to 0.001, where $\tau$ and Z are age and metallicity.
The distance modulus was explored in the range from 15 to 18 in steps of 0.01.
For each parameter combination we compute the probability that a given ridge
line point was drawn from the isochrone at its minimum distance position to that
given point. The probability is computed assuming a normal distribution with a
$1\sigma$ dispersion as indicated by the error bars. All ridge line points are
given the same weight. An obvious improvement would be to weight these points by
a mass function (MF). However, the MF of a stream is likely very different from
an initial MF (IMF) and can vary along the stream itself \citep{Koch04}. For
this reason, we leave the study of the MF of the stream to future works with
deeper photometry and more accurate membership probabilities.

We calculate a likelihood function for our model by multiplying the individual
probabilities of each ridge line point. We define the best model as the one
maximizing the likelihood and the parameters' uncertainties are derived using
the profile likelihood technique \citep[e.g.][]{Rolke05}. To estimate the 90\%
confidence interval of each fitter parameter we find the value of that parameter
where the log-likelihood (maximized with respect to the other parameters)
decreases by 2.71/2 from its maximum value. We find that the stream population
is well described by a model with $(m-M) = 16.21\pm0.11$ (or $d_{\odot} = 17.5
\pm 0.9$ kpc), $\log_{10}(\tau/yr) = 10.06 \pm0.02$ (or $\tau = 11.5 \pm 0.5$
Gyr), and $Z < 0.0004$ (or ${\rm [Fe/H]} < -1.6$). The best-fit model is shown
in \autoref{cmd_hess} as the solid black line. Notice that the lowest
metallicity available in our model grid is still consistent with the stream CMD,
thus we are only able to define a upper bound for the metallicity. We summarize
the stream parameters in \autoref{tab:pars}.

\begin{figure*}
\centering
\includegraphics[width=0.95\textwidth]{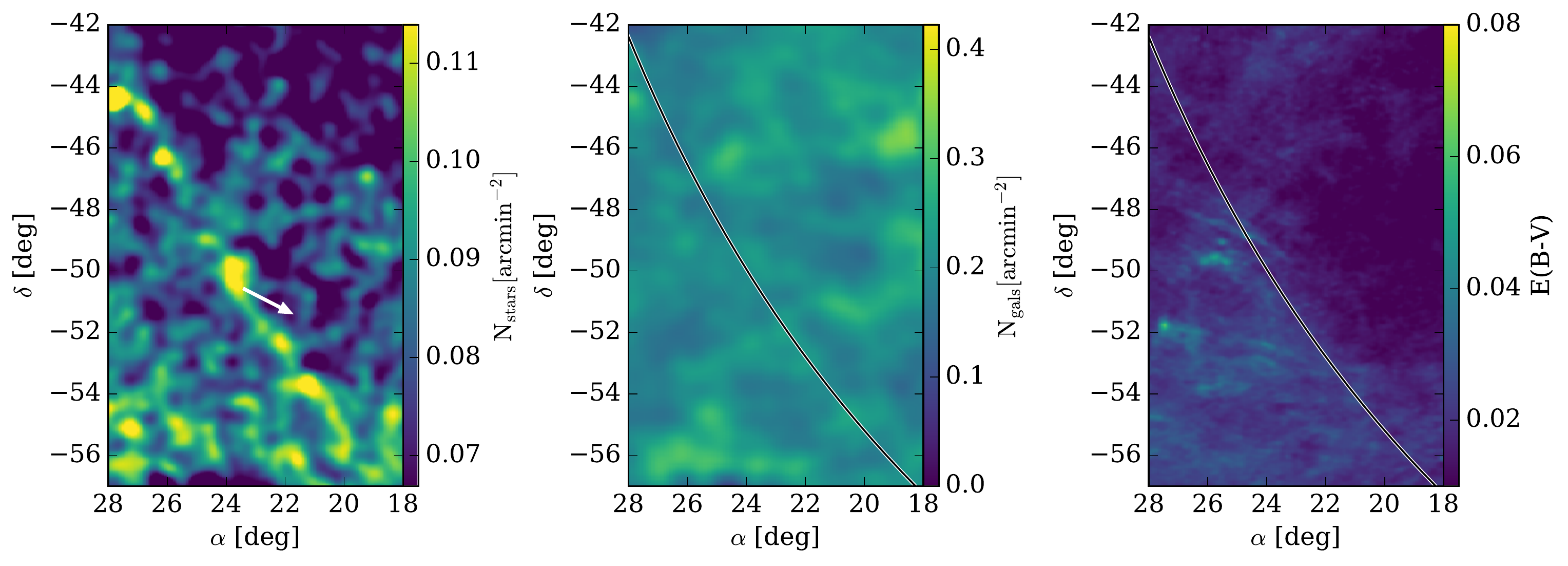} 
\caption{\emph{Left panel:} density map after applying the color-magnitude
    selection shown in Figure 2. The white arrow points towards the MW center.
    \emph{center panel:} density map using the same filter as the previous panel
    but built using sources classified as galaxies.  \emph{Right panel:} E(B-V)
    color excess map from \citet{SFD} and \citet{2011ApJ...737..103S}. On the
    last two maps the solid line shows the position of the stream as seen in the
stellar density map. All maps use a pixel size of $4.5\arcmin\times2.8\arcmin$.}
\label{maps}
\end{figure*}

We use the best-fit model to define a region in the color magnitude diagram
(CMD) where stream stars are more likely to be. This region is shown as the
dashed line in \autoref{cmd_hess}. The locus shown in the figure was defined by
color shifting the best-fit model by twice the typical color error at each
magnitude value. We also consider that the color uncertainty at magnitudes
brighter than $g=21$ is constant and equal to 0.03.

In \autoref{maps} we show the density map for stars (left panel) built using
the color selection described above but only for stars with $20 < g < 23$. The
best-fit model describes the stream population in the full domain of colors and
magnitudes observed, however we find that the HB, Sub-giant
Branch (SGB), and Red Giant Branch (RBG) are very sparsely populated. Including
these stars in the CMD selection adds more noise than signal to our density
maps. In the same figure, we also show the density map, using the same selection
as before, but for sources classified as galaxies (center panel).  On the
rightmost panel we show a reddening map from \citet{SFD}. On the last two panels
the solid line shows the position of the stream. We notice no obvious features in
the galaxy or reddening distribution that could mimic the presence of the
stream.

\subsection{Possible progenitors}

In order to investigate any possible progenitors for the new stream we assume
that streams are approximate tracers of the progenitor's orbit \citep{Bovy14}.
We also exploit the property of spherically symmetric potentials in which orbits
should be confined to a plane containing the center of such a potential
\citep{BT08}. The same is approximately true for axisymmetric potentials
\citep{Johnston96}. For more complex potentials this assumption only holds close
to the progenitor itself.  There is evidence that the MW potential, at least in
its inner parts, is well approximated by an axisymmetric potential
\citep[see][]{Kuepper15}. The Phoenix stream lies at 18.4 kpc from the Galactic
center where the MW potential should be reasonably spherical
\citep{Kuepper15, Deason11, Bell08}. Under the assumption that the stream formed
through the interaction with the MW potential only, we expect that it should be
confined to a plane passing through the center of the Galaxy. When observed from
the center of the MW this plane is described by a great circle.

To define such a plane we choose three anchor points along the stream. These
points are defined by their Galactic coordinates and the heliocentric distance
to the stream. We find the plane that contains the anchor points and the MW
center. And finally we find the circle oriented the same way as the plane that
intersects all anchor points. In order to intersect all three anchor points we
must apply corrections to their heliocentric distances, which were so far
considered all identical.

We find that a heliocentric distance gradient of $\sim$ 1 kpc is necessary for a
circular orbit to intersect all three anchor points. In order to check this
possible distance gradient we build two separate decontaminated Hess diagrams
following the same procedure outlined in \autoref{sec:res}: one using stars
north of $\delta = -56^{\circ}$ and the other using stars South of this same
declination value. Using the ridge line of each of these new Hess diagrams we
proceed with the fitting process, however this time we keep the metallicity and
age fixed at the best-fit values found previously using the full stream length.
We find that the best-fit distance modulus for the North part of the stream is
$16.19\pm0.12$, and $16.35\pm0.12$ in the South. This is consistent with, but
does not require, the $\sim$1 kpc gradient required for a circular orbit. We
conclude that a distance gradient cannot be ruled out for the stream. Detailed
spectroscopic observations must be used to isolate stream member stars based on
radial velocities and chemical composition and confirm this scenario. For the
purpose of looking for possible progenitors close to the stream we will assume a
circular orbit.

\begin{figure}
\centering
\includegraphics[width=0.45\textwidth]{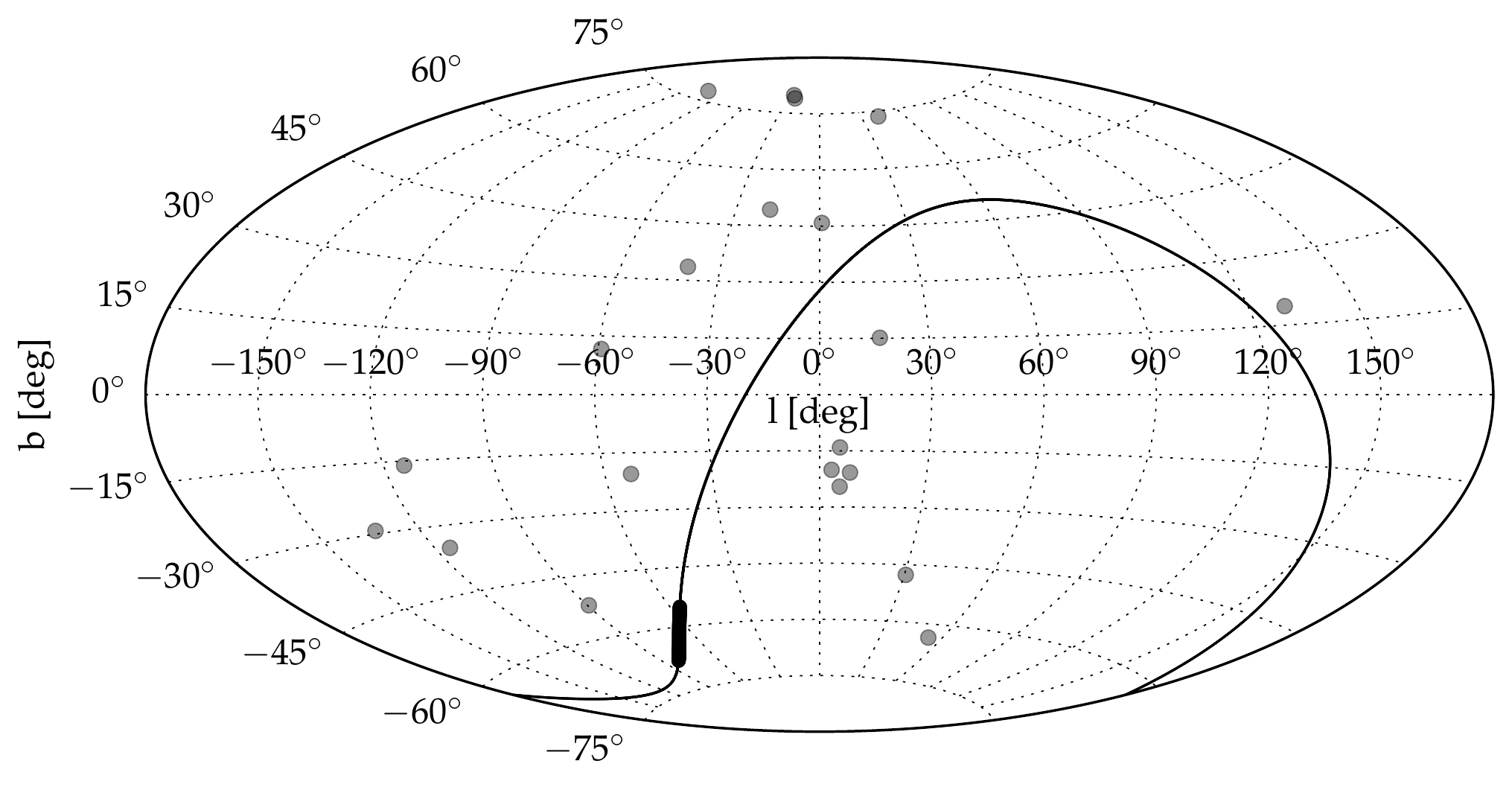}
\caption{Aitoff projection in Galactic coordinates. The gray circles show
GCs with Galactocentric distances between 15 and 25 kpc. The solid
line shows the great circle that best fits the Phoenix stream, the portion of
the stream observed is highlighted with a broader line.}
\label{allsky}
\end{figure}

Using the best-fit circular orbit described above, we look for possible known
GCs that are not in the Y1A1 footprint that could be progenitors. This approach
does not explore other kinds of orbits, which are more likely, however it
provides a useful approach to search for progenitors in the close vicinity of
the stream. In \autoref{allsky} we show the best-fit great circle in an all-sky
Aitoff Galactic projection. We also show known GCs with Galactocentric distances
between 15 and 25 kpc, distances consistent with the stream distance. We find
that no GC is consistent with this stream under the assumption of a circular
orbit.

Another possible scenario is that the progenitor has completely dissolved and
only its remains are visible along the stream. To investigate possible
progenitors for this stream we first look for overdensities on the stream. We
start by creating a reference frame where the horizontal axis is oriented along
the stream, similar to what has been adopted by \citet{Majewski03} for the
Sagittarius stream. To create such a reference frame we use two Euler angles
that define two consecutive rotations ($\phi, \theta$). The angles were
determined by finding the plane that intersects the anchor stream points in
Equatorial coordinates. The new reference frame has an azimuthal component
($\Lambda$) with an arbitrary origin and defined in the range [0, 2$\pi$), and
an elevation component ($\beta$) defined in the range
[$-\frac{\pi}{2},\,\frac{\pi}{2}$]. The values of $\phi$ and $\theta$ adopted
are -29$\fdg$698 and 72$\fdg$247 respectively.

\begin{figure*}
\centering
\includegraphics[width=0.85\textwidth]{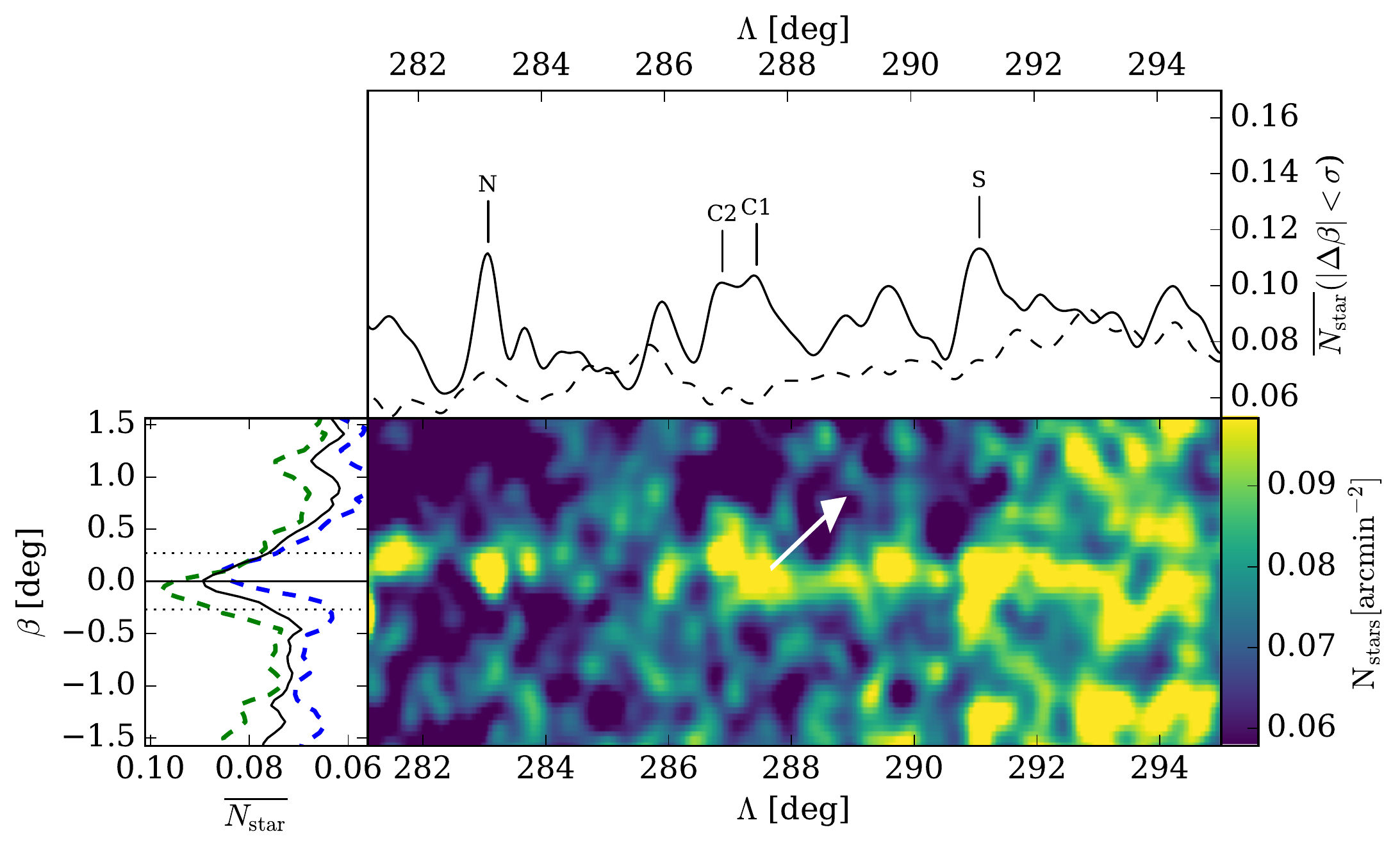}
\caption{\emph{Bottom right panel:} density map in the stream's coordinate
    system built using color-magnitude selected stars. The white arrow points
    towards the MW center. The pixel size is the same as in
    \autoref{colorselect}. The dashed lines shown in Figure 2 encompass the
    region of the CMD where stars were selected from. The \emph{left panel}
    shows the average density map value in projection onto the $\beta$ axis, the
    solid horizontal line marks the position of the stream centroid and the
    dotted lines shows the $\pm1\sigma$ limits. The green (blue) dashed line
    shows the same kind of $\beta$ projection but for the South (North) part of
    the stream.  The \emph{top panel} shows the average density map value in
    projection along the $\Lambda$ axis. This latter is built using only stars
    that are within $\pm1\sigma$ of the centroid of the stream. Vertical solid
    lines mark the position of candidate progenitors. In this panel the dashed
line shows the typical background contribution, computed using stars within $\pm
1 \sigma$ of a center line offset $5\sigma$ from the stream's original center
line.}
\label{profile}
\end{figure*}

In \autoref{profile} we show the density map of color-selected stars in the
coordinate system described above. We also show the average density map in
projection onto the $\beta$ (\emph{left panel}) and $\Lambda$ (\emph{top panel})
axes. The $\Lambda$ projection is shown for stars that lie within $\sigma =
0^{\circ}.18$ from the stream centroid in $\beta$, where $\sigma$ is the
standard deviation with respect to the stream median line. The stream median
line is defined in the stream coordinate system as the peak in the $\beta$
projected density. We also show the $\beta$ projection for stars south (green
dashed line) and north (blue dashed line) of the central overdensity. We notice
an offset of $\sim 0^{\circ}.14$ in $\beta$ between the North and South portion
of the stream.

\begin{table}
\caption{Phoenix stream parameter summary}
\label{tab:pars}
\begin{tabular}{lcccl}
\hline
Name & Value & Unit & Description \\
\hline
$\tau$       &  $11.5\pm0.5$                 & Gyr &  Age         \\
${\rm [Fe/H]}$       &  $< -1.6$   & dex &  Metallicity \\
$d_{\odot}$  &  $17.5\pm0.9$                 & kpc &  Heliocentric distance \\
$d_{GC}$     &  $18.4\pm0.9$                 & kpc &  Galactocentric distance* \\
$\sigma$     &  54                           & pc  &  Stream width$^\dagger$ \\
$(\alpha, \delta)_{start}$ & (20, -57)        & deg &  Stream begin point \\
$(\alpha, \delta)_{end}$   &   (27, -45)        & deg &  Stream end point \\
$\Theta$       &   8.1                         & deg &  Stream length \\
\hline
\end{tabular} \\
* Computed using $R_{\odot}$ = 8.3 kpc; $\dagger$ assuming a Gaussian profile.
\medskip
\end{table}

The stream width depends on two sets of factors. The first and more obvious is
the progenitor's size and velocity dispersion. The second is the shape of the
gravitational potential. For instance, triaxial potentials tend to increase the
fanning of stream stars significantly \citep{apw16, Pearson15}. However, there is
evidence that the inner halo of the Galaxy is relatively spherical
\citep{Kuepper15} out to $\sim 20$ kpc, thus allowing us to assume the stream
width maps only to the progenitor size and velocity dispersion. Using the stream
coordinate system we determine that it has an on-sky width of $\sigma =
0^{\circ}.18$, which translates to $\sim 54$ pc at its distance. Typically, 50
pc is consistent with the tidal radius of the MW halo GCs. The fact that
the stream forms a thin coherent structure several kpc long makes it plausible
that the progenitor was in fact a GC.

From the density projected in the $\Lambda$ coordinate we explore the presence
of overdensities (ODs) as possible progenitors. We label these ODs in the top
panel of \autoref{profile}. First we call attention to C1 and C2, which have a
slight offset with respect to each other in the $\beta$ direction, showing hints
of trailing and leading tail misalignment with respect to the orbit
\citep{Bovy14}. Apart from C1 and C2 we find two other peaks, one to the north
(N) and another to the south (S) of the central ODs. These ODs stand out when
compared to the typical background counts (dashed line in \autoref{profile}).
We compute the typical local background noise at the position of the north
(south) OD by taking the standard deviation of the background counts in the
north (south) portion of the stream. We find that both ODs (N \& S) peak
densities stand out more than 4$\sigma$ with respect to the background.  The
significance values are listed on Table 2.

The fact that ODs N and S are approximately equally separated from the central
OD could point to epicyclic overdensities such as the ones reported in
\citet{Kuepper15} for Palomar 5. \autoref{tabods} summarizes the positions
and angular separation of the overdensities with regard to the central peaks.
From this table we observe that ODs to the north are systematically at higher
$\beta$ than those in the South, hinting to the misalignment mentioned above.

Using the misalignment of the Northern and Southern portions of the stream
(hinted in \autoref{profile}), we may infer from geometrical considerations
alone that its Northern part is closer to the MW center, hence being formed by
stars that leave the progenitor through the inner part of its orbit, forming a
leading tail. By construction, the South portion forms the trailing tail. From
this argument, we conclude that the stream is moving from south to north.

All overdensities were visually inspected on the coadded images and catalog,
however we could not identify any stellar system (e.g. globular cluster) that
might have given origin to the stream. This result is very puzzling, especially
if the scenario described above is to be confirmed. The fact that no progenitor
is found, but classic signs of cold tail formation are observed could indicate
that a progenitor was fully disrupted very recently.

\section{Conclusions}

We report the discovery of a stellar stream in the Southern hemisphere. Through
the visual fit of stellar evolution models we found that this stream is
comprised of an old (11.5$\pm$0.5 Gyr) metal-poor (${\rm[Fe/H]} < -1.6$)
population that is 17.5 kpc away from the Sun and 18.4 kpc from the Galactic
center. Though close in projection, the Phoenix stream is not related to the
Phoenix galaxy, which lies at 440 kpc from the Sun \citep{2004AJ....127.2031K}.

\begin{table}
    \scriptsize
\caption{Overdensity positions}
\label{tabods}
\begin{tabular}{lcccccccc}
\hline
OD & RA & Dec & $l$&$b$ & $\Lambda$&$\beta$ & $\Delta\overline{\rm C1C2}$ & $\sigma$\\
   & deg & deg & deg & deg & deg& deg & deg & -- \\
\hline
C1  &  23.75&  -49.89 & 285.68& -65.76 & 287.50& 0.00 &   --  & 5.1  \\
C2  &  23.77&  -50.40 & 286.16& -65.28 & 286.94& 0.22 &   --  & 5.0  \\
N  &  26.15 & -46.34 &  277.74& -68.10 & 283.16& -0.09 & 4.10 & 7.1 \\
S  &  21.23 & -53.70 &  292.10& -62.72 & 291.11& 0.03  & 4.17 & 5.0 \\
\hline
\end{tabular}
\end{table}

Through the extrapolation of the stream outside the Y1A1 footprint, we found no
known GC that could be its progenitor; however more eccentric and/or non-planar
orbits were not considered. 

We also investigate the distribution of overdensities along the stream in search
for a progenitor. We find that none of the ODs has any obvious stellar
overdensity associated with it when coadded images were inspected. We find that
the ODs with high significance display a symmetric pattern with respect to a
central overdensity.  This central overdensity shows some hints of misalignment
perpendicular to the orbit direction, which could indicate the position of the
progenitor.

A diffuse stellar overdensity has recently been found in the DES data which
nearly overlaps with the Phoenix stream \citep{eriphe}. This overdensity
(EriPhe) was previously hinted at by \citet{2014MNRAS.445.2971C} as an anomalous
background population close to NGC 1261. EriPhe and NGC 1261 share a similar
heliocentric distance as the Phoenix stream. Using \textsc{galpy} \citep{Bovy15}
and literature proper motions for NGC 1261 \citep{Dambis06}, we integrate the
cluster orbit and find that it roughly aligns with the Phoenix stream and that
its motion is retrograde with respect to the Solar motion.  The close proximity
of NGC 1261 and the orbit alignment with the stream may suggest that they could
share a common origin with the EriPhe overdensity. However, the stream appears
not to be in a retrograde orbit, favouring a scenario where the stream is
independent from EriPhe or NGC 1261.  For an extended discussion about this
scenario we refer to \citet{eriphe}.

\section*{Acknowledgments}

We thank Carl Grillmair for pointing out that our original derivation of
the direction of motion of the stream was incorrect, and that the stream is
actually moving in a prograde direction about the Galaxy.

Funding for the DES Projects has been provided by the U.S. Department of
Energy, the U.S. National Science Foundation, the Ministry of Science and
Education of Spain, the Science and Technology Facilities Council of the United
Kingdom, the Higher Education Funding Council for England, the National Center
for Supercomputing Applications at the University of Illinois at
Urbana-Champaign, the Kavli Institute of Cosmological Physics at the University
of Chicago, the Center for Cosmology and Astro-Particle Physics at the Ohio
State University, the Mitchell Institute for Fundamental Physics and Astronomy
at Texas A\&M University, Financiadora de Estudos e Projetos, Funda{\c c}{\~a}o
Carlos Chagas Filho de Amparo {\`a} Pesquisa do Estado do Rio de Janeiro,
Conselho Nacional de Desenvolvimento Cient{\'i}fico e Tecnol{\'o}gico and the
Minist{\'e}rio da Ci{\^e}ncia, Tecnologia e Inova{\c c}{\~a}o, the Deutsche
Forschungsgemeinschaft and the Collaborating Institutions in the Dark Energy
Survey.

The Collaborating Institutions are Argonne National Laboratory, the University
of California at Santa Cruz, the University of Cambridge, Centro de
Investigaciones Energ{\'e}ticas, Medioambientales y Tecnol{\'o}gicas-Madrid,
the University of Chicago, University College London, the DES-Brazil
Consortium, the University of Edinburgh, the Eidgen{\"o}ssische Technische
Hochschule (ETH) Z{\"u}rich, Fermi National Accelerator Laboratory, the
University of Illinois at Urbana-Champaign, the Institut de Ci{\`e}ncies de
l'Espai (IEEC/CSIC), the Institut de F{\'i}sica d'Altes Energies, Lawrence
Berkeley National Laboratory, the Ludwig-Maximilians Universit{\"a}t
M{\"u}nchen and the associated Excellence Cluster Universe, the University of
Michigan, the National Optical Astronomy Observatory, the University of
Nottingham, The Ohio State University, the University of Pennsylvania, the
University of Portsmouth, SLAC National Accelerator Laboratory, Stanford
University, the University of Sussex, and Texas A\&M University.

The DES data management system is supported by the National Science Foundation
under Grant Number AST-1138766.  The DES participants from Spanish institutions
are partially supported by MINECO under grants AYA2012-39559, ESP2013-48274,
FPA2013-47986, and Centro de Excelencia Severo Ochoa SEV-2012-0234, some of
which include ERDF funds from the European Union. Research leading to these
results has received funding from the European Research Council under the
European Union’s Seventh Framework Programme (FP7/2007-2013) including ERC grant
agreements 240672, 291329, and 306478.

This research made use of Astropy, a community-developed core Python package for
Astronomy \citep{astropy}.

EBa acknowledges financial support from the European Research Council
(ERC-StG-335936, CLUSTERS).

This paper has gone through internal review by the DES collaboration.

%

\bibliographystyle{apj}            
\bibliography{refs}                 

\label{lastpage}

\end{document}